\documentclass[12pt,preprint]{aastex}

\newcommand{\ba}{\begin{eqnarray}}
\newcommand{\ea}{\end{eqnarray}}

\begin{document}

\title{GeV and higher energy photon interactions in gamma-ray burst
fireballs and surroundings}

\author{Soebur Razzaque$^1$, Peter M\'esz\'aros$^{1,2,3}$ and Bing
Zhang$^1$} \affil{$^1$Astronomy \& Astrophysics Department,
Pennsylvania State University, University Park, PA 16802 \\
$^2$Physics Department, Pennsylvania State University, University
Park, PA 16802 \\ $^3$Institute for Advanced Study, Einstein Drive,
Princeton, NJ 08540}

\begin{abstract}
We have calculated the opacities and secondary production mechanisms
of high energy photons arising in gamma-ray burst internal shocks,
using exact cross-sections for the relevant processes. We find that
for reasonable choices of parameters, photons in the range of 10's to
100's of GeV may be emitted in the prompt phase. Photons above this
range are subject to electron-positron pair production with fireball
photons and would be absent from the spectrum escaping the gamma-ray
burst. We find that, in such cases, the fireball becomes optically
thin again at ultra-high energies ($\gtrsim$ PeV). On the other hand,
for sufficiently large fireball bulk Lorentz factors, the fireball is
optically thin at all energies. Both for $\gamma\gamma$ self-absorbed
and optically thin cases, the escaping high energy photons can
interact with infra-red and microwave background photons to produce
delayed secondary photons in the GeV-TeV range.  These may be
observable with GLAST, or at low redshifts with ground-based air
Cherenkov telescopes. Detection of the primary prompt spectrum
constrains the bulk Lorentz factor, while detection of delayed
secondary gamma-rays would provide a consistency check for the primary
spectrum and the bulk Lorentz factor as well as constraints on the
intergalactic magnetic field strength.
\end{abstract}


\section{Introduction}

GeV photons have been detected from a number of gamma-ray bursts (GRB)
\citep{dingus03} with the EGRET detector on the Compton Gamma Ray
Observatory. Photons at energies in this range and higher can make
pairs via $\gamma\gamma \to {\rm e}^+ {\rm e}^-$ interactions in the
GRB fireball, the optical depth being dependent on the bulk Lorentz
factor $\Gamma_b$, so spectral observations at this energy can provide
useful constraints \citep{baring94, lithwick01} on this key ingredient
of GRB models.  In addition, high energy photons which escape
$\gamma\gamma$ absorption in the source can still suffer such
interactions far from the source but before reaching the observer,
against cosmic infrared background (CIB) and/or cosmic microwave
background (CMB) photons. The electron-positron pairs thus produced
can inverse-Compton scatter again on the most numerous CMB photons,
giving rise to a delayed secondary spectrum \citep{plaga95, dai02a},
which ranges typically up to 100's of GeV.

In this paper we investigate, using the exact cross sections, the high
energy interactions of photons produced in a simple GRB internal shock
model, as a function of the luminosity, the typical variability
timescale and the bulk Lorentz factor. We approximate the original GRB
spectrum through an observationally motivated Band function
\citep{band93}, and calculate the emerging spectrum as a function of
the burst parameters. This spectrum is then subjected to interactions
with the CIB and CMB, and the shape of the secondary spectrum and its
time delay is calculated assuming typical cosmological distances
$z\sim 1$ and various values for the intergalactic magnetic field
strength. The pair-production cut-off of the primary spectrum and the
delayed secondary spectrum fall, for typical GRB parameters, in the
GLAST \citep{gehrels99} energy range, and for a range of intergalactic
magnetic field strengths the latter should be detectable by GLAST as
well as by the newer generation of air-Cherenkov telescopes
\citep{weekes01}.

\section{GRB internal shock model}

Denoting with $U$ the GRB fireball total energy density, after the
initial expansion phase the kinetic energy is carried by the baryons
since $m_p \gg m_e$, where $m_p$ and $m_e$ are baryon and lepton
masses respectively. Thus, $U \sim U_p \sim 4 n'_p m_p c^2
\Gamma_b^2$, where $n'_p$ is the total baryon number density in the
comoving fireball frame and $\Gamma_b$ is the bulk Lorentz factor of
the fireball in the observer's frame.\footnote{We use primed variables
in the comoving frame and un-primed variables in the laborotary or
observer's frame} Later on, a significant fraction of the fireball's
kinetic energy is transferred to leptons, in internal shocks which
randomize the relative bulk motion between different portions of the
expanding fireball, as well as external shocks when the fireball is
decelerated by the external medium. Here we concentrate on the
internal shocks, which are thought to be responsible for the usual
prompt MeV emission. The fraction of energy transferred to the leptons
is given by a parameter $\varepsilon_e = U_e/U \gg m_e/m_p$.  Charged
particles, $p$ and $e^+e^-$, are accelerated in the shocks by
interaction with magnetic fields within the GRB fireball.  High energy
protons and electrons cool mostly due to synchrotron radiation, but
the proton cooling time is much longer than that of the electrons, and
the prompt $\gamma$-ray emission is most likely due to electron
synchrotron emission, whose cooling time in the prompt phase (tens of
seconds after the trigger) is generally shorter than the dynamical or
expansion time.

In this so-called fast cooling case, valid in the internal shocks,
most of the electron energy is converted into $\gamma$-rays: $U_e
\approx U_{\gamma} = L_{\gamma,{\rm iso}}/(4\pi r_{\rm sh}^2 c)$.
Here $r_{\rm sh} = 2c\delta t \Gamma_b^2$ is the shock radius, $\delta
t$ is the observed variability time scale, and $L_{\gamma,{\rm iso}}$
is the isotropic-equivalent $\gamma$-ray luminosity in the observer's
frame during the prompt $\gamma$-ray emission phase.  Neglecting
$e^{\pm}$ pair formation in the shocks, the total volume number
density of leptons and baryons in the fireball (which are the same due
to charge neutrality) can be calculated, assuming $U_p \approx
U_{\gamma}/\varepsilon_e$, as
\ba n'_e \approx n'_p = \frac{L_{\gamma,{\rm iso}}/\varepsilon_e}{4\pi
r_{\rm sh}^2 \Gamma_b^2 {\bar \gamma_p} m_p c^3}
\label{co-part-den} \ea
in the comoving frame. Here ${\bar \gamma_p} \sim 1$ is the comoving
random Lorentz factor acquired by protons in the internal shocks. The
magnetic field energy density $U_B = B^2/8\pi$ can be estimated as
\ba B' = \sqrt{\frac{2 \varepsilon_B L_{\gamma,{\rm iso}}}{
\varepsilon_e r_{\rm sh}^2 c \Gamma_b^2 }},
\label{co-bfield} \ea
so in the comoving frame $\varepsilon_B = U_B/U$ is the fraction of
the shocked fireball energy that goes into magnetic energy.

The process of Fermi acceleration in the shocks leads to a
differential electron and proton volume number density distribution
which is a power law in energy,
\ba \frac{dN'_{e,p}}{d\gamma'_{e,p}} = \frac{(p-1) n'_{e,p}}
{\gamma'_{e,p;{\rm min}}} \left( \frac{\gamma'_{e,p}}
{\gamma'_{e,p;{\rm min}}} \right)^{-p} ; \;\; \gamma'_{e,p} >
\gamma'_{e,p;{\rm min}}
\label{ep-dist} \ea
where
\ba \gamma'_{e,{\rm min}} \approx \left( \frac{m_p}{m_e} \right)
\varepsilon_e \gamma'_{p,{\rm min}} \; ; \; \gamma'_{p,{\rm min}}
\approx 1
\label{min-boost} \ea
are the minimum Lorentz factors for electrons and protons respectively
in the comoving frame. The maximum Lorentz factors are obtained by
equating the acceleration time ($t'_{j,{\rm acc}} = A m_{j}
\gamma_{j}/eB'$) of a particle $j$ to the shorter of the synchrotron
cooling time
\ba t'_{j;syn} = \frac{6\pi m_{j}^3} {\sigma_{\rm Th} \gamma_{j} m_e^2
B^{'2}}
\label{syn-cooltime} \ea
and the dynamic time ($\delta t \Gamma_b$) in the comoving frame.
Electrons cool faster, by synchrotron radiation, than the dynamic time
and the maximum electron Lorentz factor is
\ba \gamma'_{e, {\rm max}} &=& \sqrt{ \frac{6\pi e }{A \sigma_{\rm Th}
B'}} \nonumber \\ &\approx& 2.2\times 10^6 A_1^{-1/2}
L_{\gamma,52}^{-1/4} \varepsilon_{e,-1}^{1/4}
\varepsilon_{B,-1}^{-1/4} \delta t_{-2}^{1/2} \Gamma_{b,2.5}^{3/2}
\label{max-eboost} \ea
in the comoving frame. Here we have chosen $L_{\gamma,{\rm iso}} =
10^{52}L_{\gamma,52}$ erg/s, $\varepsilon_e = 10^{-1}
\varepsilon_{e,-1}$, $\varepsilon_B = 10^{-1} \varepsilon_{B,-1}$,
$\delta t = 10^{-2} \delta t_{-2}$ s, $A=10A_1$ and $\Gamma_{b} =
10^{2.5} \Gamma_{b,2.5}$.  Note that the inverse Compton (IC) cooling
time of the electrons is proportional to the synchrotron cooling time
in equation (\ref{syn-cooltime}) and the proprtionality constant, in
the fast cooling case, is given by $Y= (-1+\sqrt{1+4
\varepsilon_e/\varepsilon_B})/2$ \citep{se01}.  For the present choice
of parameters, $Y\approx 0.6$ and we have ignored the IC cooling of
the electrons compared to the synchrotron cooling while calculating
$\gamma'_{e, {\rm max}}$.  Our choice of the value of $\varepsilon_B$
is motivated by the assumption that the observed peak $\gamma$-ray
energy is due to electron synchrotron radiation and only a small
fraction of the total $\gamma$-ray power is radiated by IC process at
small internal shock radii \citep{daigne98, derishev01, baring04}.  A
recent study by \cite{zkm03} also suggest that the fireball value of
$\varepsilon_B$ should be 10-100 times larger than the afterglow fit
of 0.01-0.001.

The observed $\gamma$-ray spectrum from the GRB prompt emission can be
approximated by the phenomenological Band function in the high energy
range as $\sim \epsilon_{\gamma}^{-\alpha}$ above the synchrotron peak
energy $\epsilon_{\gamma,{\rm pk}}$, which is typically a few hundred
keV. For observed cases of $\alpha >2$, most of the $\gamma$-ray
energy is concentrated near $\epsilon_{\gamma,{\rm pk}}$ and one can
roughly estimate the peak volume number density of photons in the
fireball as
\ba n'_{\gamma} = \frac{L_{\gamma,{\rm iso}}} {4\pi r_{\rm sh}^{2}c
\Gamma_b \epsilon_{\gamma,{\rm pk}}}
\label{co-phot-den} \ea
in the comoving frame. Below the synchrotron peak energy the Band
spectrum is $\sim \epsilon_{\gamma}^{-1}$ and the total number density
of photons in equation (\ref{co-phot-den}) increases only
logarithmically.

Low energy photons are expected to be absorbed by fireball electrons
in the presence of a magnetic field by the synchrotron self-absorption
mechanism below a photon energy of
\ba \epsilon_{\gamma, {\rm ssa}} &=& \Gamma_b \left[ \frac{\pi
(-1)}{2} \Gamma\left( \frac{p}{4} + \frac{11}{6} \right) \Gamma \left(
\frac{p}{4} + \frac{1}{6} \right) \gamma_{e,{\rm min}}^{p-1} n'_e
r'_{sh} \right]^{\frac{2}{p+4}} \left[ 3^{p+1} \frac{q^{p+6} (B'
\sin\theta)^{p+2} }{m_e^{p+4}} \right]^{\frac{1}{p+4}} \nonumber \\
&\approx & \cases{ 11 L_{\gamma, {\rm 52}}^{2/3}
\varepsilon_{e,-1}^{-2/3} \varepsilon_{B,-1}^{1/3} \delta t_{-2}^{-1}
\Gamma_{b,2.5}^{-8/3}~{\rm keV};~p=2 \cr 13 L_{\gamma, {\rm
52}}^{17/26} \varepsilon_{e,-1}^{-17/26} \varepsilon_{B,-1}^{9/26}
\delta t_{-2}^{-1} \Gamma_{b,2.5}^{-34/13}~{\rm keV};~p=2.5 \cr 14
L_{\gamma, {\rm 52}}^{9/14} \varepsilon_{e,-1}^{-9/14}
\varepsilon_{B,-1}^{5/14} \delta t_{-2}^{-1}
\Gamma_{b,2.5}^{-18/7}~{\rm keV};~p=3 }
\label{ssa-energy} \ea
in the observer's frame. Here we have estimated $\epsilon_{\gamma,
{\rm ssa}}$ for power law indices in equation (\ref{ep-dist}) of $p =
2, 2.5, 3$ (top to bottom in the equation above). Observed GRB
spectra, fitted by a Band function, do not correspond in a
straightforward manner to a theoretical synchrotron spectrum from a
one-zone region with some power-law index $p$, especially in the
fast-cooling regime. However, the usual Band function values can arise
as a result of the superposition of many shocks, each giving rise to a
power law energy distribution with index $2\lesssim p \lesssim 3$. In
our calculation, we assume the nominal Band fit values for the
resulting photon spectrum, and we use a nominal reference value for
the absorption energy of $\epsilon_{\gamma, {\rm ssa}} =
10\epsilon_{\gamma,1}$ keV, which is more than an order of magnitude
below the observed synchrotron peak energy of $\epsilon_{\gamma,{\rm
pk}} = 500\epsilon_{\gamma,2.7}$ keV.

\section{High energy photon interactions and internal attenuation}

Observation of high energy photons from the GRB fireball is mainly
limited by the opacity of two photon annihilation into an
electron-positron pair ($e^{\pm}$). The cross-section for
$\gamma\gamma \rightarrow e^{\pm}$ in the non-relativistic (NR) and
extremely relativistic (ER) cases is given by \citet{jauch} as
\ba \sigma_{\gamma\gamma}^{\rm NR} (\omega) &=& \frac{3}{8}
\sigma_{\rm Th} \sqrt{1- \frac{m_e^2 c^4}{\omega^2}} \nonumber \\
\sigma_{\gamma\gamma}^{\rm ER} (\omega) &=& \frac{3}{8} \sigma_{\rm
Th} \frac{m_e^2 c^4}{\omega^2} \left( {\rm ln} \frac{2\omega}{m_e c^2}
-1\right)
\label{pair-cross} \ea
where $\omega$ is the photon energy in the center of mass frame.  The
observed photon energy ($E_{\gamma}$) above the threshold for pair
production with photons at an energy $\epsilon_{\gamma}$ must satisfy
the condition $\omega > m_e c^2$ or $E_{\gamma} \gtrsim E_{\gamma,{\rm
th}} = m_e^2 c^4 \Gamma_b^2 /2\epsilon_{\gamma}$.  Observation of
photons of a maximum energy $\epsilon_{\gamma,{\rm max}} >
E_{\gamma,{\rm th}}$ may therefore be used to put a limit on
$\Gamma_b$ {\it if} the corresponding optical depth is unity.  Photons
below $E_{\gamma,{\rm th}}$ may escape from the fireball (provided no
other interaction becomes important) even though they might interact
with photons from the high energy tail of the Band function, but the
relevant optical depth is less than unity because of a much lower
photon density, at that energy, than in equation (\ref{co-phot-den}).

For a given $\Gamma_b$ and $\delta t$, the lower limit threshold of
the $\gamma\gamma$ absorption energy range within the source can be
found from the pair production threshold energy with peak synchrotron
photons of energy $\epsilon_{\gamma,{\rm pk}}$ as
\ba E_{\gamma,{\rm pk,th}} = \frac{m_e^2c^4 \Gamma_b^2}{2
\epsilon_{\gamma,{\rm pk}}} \approx 26 \Gamma_{b,2.5}^2
\epsilon_{\gamma,2.7}^{-1}~{\rm GeV}
\label{egampair-thresh} \ea
in the observer's frame, for the present choice of parameters. The
total volume number density of photons and hence the $\gamma\gamma$
optical depth in the fireball increases slowly as the photon energy
decreases from $\epsilon_{\gamma,{\rm pk}}$ to $\epsilon_{\gamma,{\rm
ssa}}$. High energy photons in the energy range capable of producing
pairs with fireball photons in the energy range $\epsilon_{\gamma,{\rm
pk}}$ - $\epsilon_{\gamma,{\rm ssa}}$ are absent in the spectrum
escaping from the source. Ultra high energy photons, however, may
escape the GRB fireball as the pair production cross-section, in the
ER limit in equation (\ref{pair-cross}), decreases with increasing
photon energy and so does the optical depth.  Ultra high energy
primary photons will appear again above this ``thinning" energy. This
energy can be found, roughly, from the optical depth corresponding to
the $\gamma\gamma$ cross-section with photons at self absorption
energy ($\epsilon_{\gamma,{\rm ssa}}$) in the ER limit in equation
(\ref{pair-cross}) as
\ba E_{\gamma, {\rm thin}} &=& \frac{3\Lambda}{64\pi}
\frac{L_{\gamma,{\rm iso}}\sigma_{\rm Th} m_e^2c^2}{\Gamma^2 \delta t
\epsilon_{\gamma,{\rm ssa}}^2} \nonumber \\ &\approx & 2 \times 10^7
\Lambda_{3} L_{\gamma,52} \Gamma_{b,2.5}^{-2} \delta t_{-2}^{-1}
\epsilon_{\gamma,1}^{-2}~{\rm GeV}
\label{egammax-thin} \ea
where $\Lambda = {\rm ln} (2\sqrt{2\epsilon_{\gamma,{\rm ssa}}
E_{\gamma}}/m_e\Gamma_b ) -1 \sim 3 \Lambda_{3}$, for the same
parameters used in equation (\ref{egammax-thin}).

\begin{figure}
\plotone{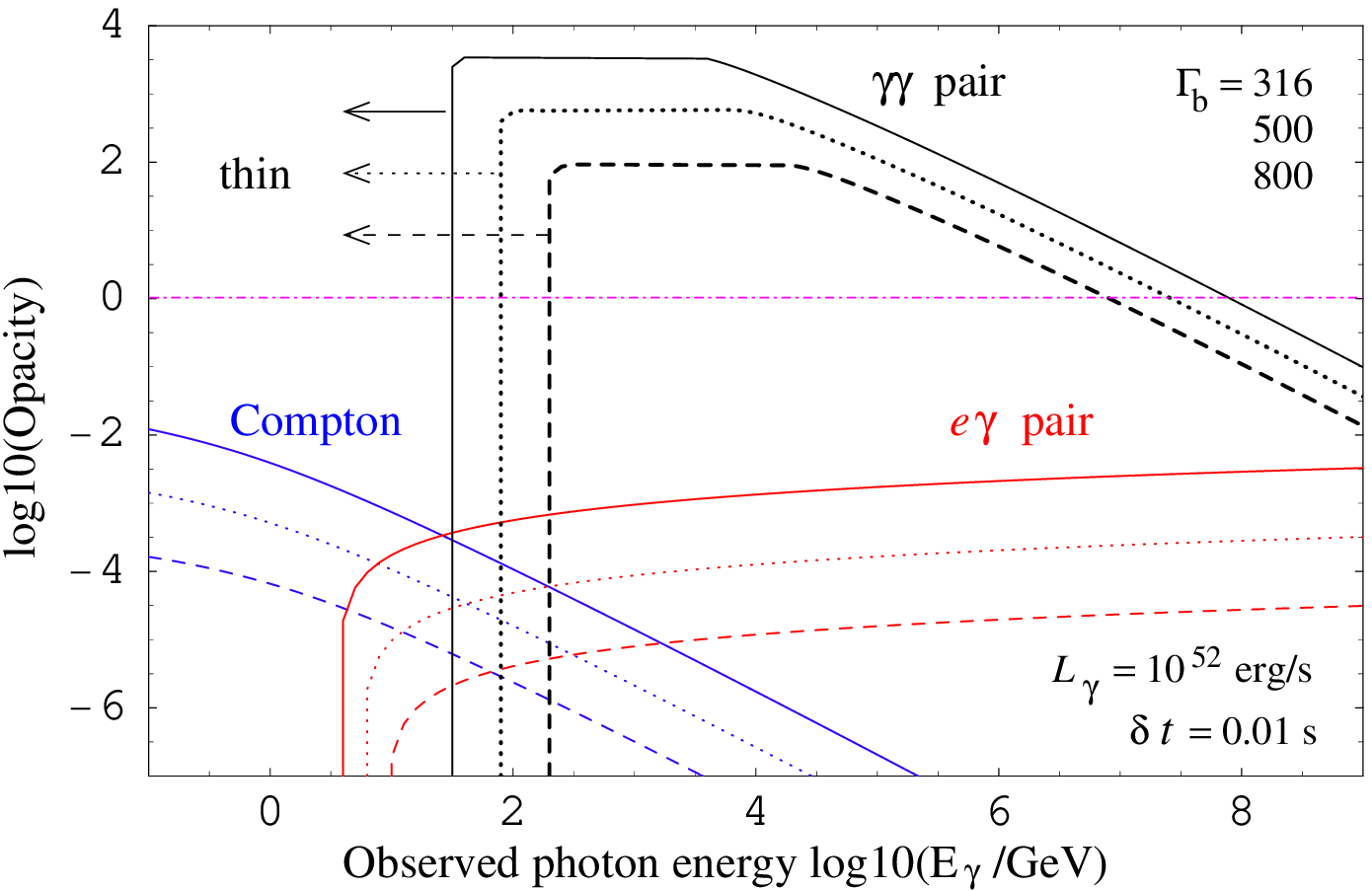} \caption{Opacity for high energy photons in the GRB
fireball, in the comoving frame, against electron Compton scattering,
$e^{\pm}$ pair production with electrons by $e\gamma$ and by
$\gamma\gamma$ interactions with fireball photons. We have plotted the
curves for $\L_{\gamma,{\rm iso}} = 10^{52}$ erg/s isotropic
$\gamma$-ray luminosity, the bulk Lorentz factor $\Gamma_b=$ 316
(solid curves), 500 (dotted curves) and 800 (dashed curves). We used a
variability time $\delta t =$ 0.01 s. Photons above $\sim 2\times
10^7$ GeV, for example, for $\Gamma_b = 316$ will escape the GRB
fireball. No photons, for the same parameters, may escape the fireball
below $\sim 2\times 10^7$ GeV down to $\sim 26$ GeV where the
$\gamma\gamma$ opacity is greater than unity. \label{wlimits52}}
\end{figure}

We calculate the high energy photon interactions with other fireball
photons and electrons using the exact cross-section formulae
\citep{jauch}. The opacities of high energy photons are plotted in
Figure \ref{wlimits52}, as a function of observed energy
($E_{\gamma}$), due to electron Compton scattering, $e^{\pm}$ pair
production with electrons ($e\gamma \rightarrow e^{\pm}$) and by
$\gamma\gamma$ interactions, dominantly, with fireball photons in the
energy range $\epsilon_{\gamma,{\rm ssa}}$ - $\epsilon_{\gamma,{\rm
pk}}$. The cross-section for the $e\gamma \rightarrow e^{\pm}$ process
is given by \citet{jauch} as
\ba \sigma_{e\gamma}^{\rm NR} (E'_{\gamma}) &=& 2.25\times 10^{-3}
\frac{\alpha}{\pi} \sigma_{\rm Th} \left( \frac{E'_{\gamma}}{m_e c^2}
-4 \right) \nonumber \\ \sigma_{e\gamma}^{\rm ER} (E'_{\gamma}) &=&
\frac{3\alpha}{8\pi} \sigma_{\rm Th} \left( \frac{28}{9} {\rm ln}
\frac{2E'_{\gamma}}{m_e c^2}- \frac{218}{27} \right)
\label{BH-cross} \ea
in the non-relativistic and extreme relativistic limits
respectively. Here $E'_{\gamma}$ is the photon energy in the comoving
frame. The flat region, in Figure \ref{wlimits52}, for the
$\gamma\gamma$ pair creation curves above $E_{\gamma,{\rm pk, th}}$
corresponds to a change in the target photon energy from
$\epsilon_{\gamma,{\rm pk}}$ to $\epsilon_{\gamma,{\rm ssa}}$ and a
slow increase of the total target photon number density $n'_{\gamma}$,
in equation (\ref{co-phot-den}), accordingly.

We have plotted the boundaries of the optically thin and thick regions
for high energy photon emission from GRB internal shocks in Figure
\ref{wphase}, for different choice of parameters $L_{\gamma}$,
$\epsilon_{\gamma,{\rm pk}}$ and $\delta t$. Note that for certain
parameter values, e.g. $L_{\gamma,{\rm iso}} = 10^{51}$ erg/s, $\delta
t = 0.1$ s and $\epsilon_{\gamma,{\rm pk}} = 0.5$ MeV; or
$L_{\gamma,{\rm iso}} = 10^{52}$ erg/s, $\delta t = 1$ s and
$\epsilon_{\gamma,{\rm pk}} = 1$ MeV, the GRB fireball is optically
thin to photons of all energies above $\Gamma_b \approx 775$ or 676
respectively.

Extra complications can arise due to increased $e^{\pm}$ pairs
produced in the fireball by $\gamma\gamma$ interactions by high energy
photons \citep{totani99, mrr01}, which can affect the photon opacities
plotted in Figure \ref{wlimits52}.  This is a complex problem,
requiring a numerical treatment \citep{peer03}. Here we use an
approximate treatment, using a pair formation rate in the comoving
frame of $c \sigma_{\gamma\gamma} (\epsilon'_{\gamma}) dN'_{\gamma,
{\rm i}}/ d\epsilon'_{\gamma}$ to estimate the total number of
$e^{\pm}$ pairs created by an incident (denoted by ``i'') high energy
photon.  The differential volume number density of incident high
energy photons capable of producing pairs, as a function of energy,
can be estimated using the Band spectrum and equation
(\ref{co-phot-den}) as
\ba \frac{dN'_{\gamma,{\rm i}}}{d\epsilon'_{\gamma}} \approx
\frac{\alpha -1}{\epsilon'_{\gamma, {\rm pk}}} \frac{n'_{\gamma}}
{(\epsilon'_{\gamma}/\epsilon'_{\gamma,{\rm pk}})^{\alpha}};~
\epsilon'_{\gamma} \gtrsim \epsilon'_{\gamma, {\rm pk,th}}.
\label{gam-energydensity} \ea
Here $\epsilon'_{\gamma, {\rm pk,th}}$ is the comoving threshold
energy for pair production which is similar to the threshold energy in
equation (\ref{egampair-thresh}) in the case of synchrotron peak
photons as targets. The target (denoted by ``t'') photon number
density, in general, may have a different energy distribution
$dN'_{\gamma, {\rm t}}/ dE'_{\gamma}$. The total volume number density
of pairs created in the fireball is thus
\ba n'_{\pm} &=& c\delta t \Gamma \int^{\epsilon'_{\gamma,{\rm
max}}}_{\epsilon'_{\gamma,{\rm th}}} d\epsilon'_{\gamma}
\frac{dN'_{\gamma,{\rm i}}}{d\epsilon'_{\gamma}}
\int^{\epsilon'_{\gamma,{\rm max}}}_{\epsilon'_{\gamma,{\rm th}}}
dE'_{\gamma} \sigma_{\gamma\gamma} \left( \sqrt{2 \epsilon'_{\gamma}
E'_{\gamma}} \right) \frac{dN'_{\gamma,{\rm t}}} {dE'_{\gamma}},
\label{tot-pairdensity-def} \ea
where $\epsilon'_{\gamma,{\rm max}} \approx \gamma'_{e, {\rm max}} m_e
c^2$ is defined in equation (\ref{max-eboost}), and
$\epsilon'_{\gamma,{\rm th}}=m_e^2 c^4/2E'_{\gamma}$. For this
estimate, we approximate the $\gamma\gamma$ cross-section as
$\sigma_{\gamma\gamma} \approx (3/8)\sigma_{\rm Th}$ at $e^{\pm}$ pair
production threshold with peak synchrotron photons
($\epsilon_{\gamma,{\rm pk}}$). This leads to a simplified expression
for the total volume number density of pairs created in the fireball
by integrating over the photon energy distribution above
$\epsilon'_{\gamma,{\rm th}}$,
\ba n'_{\pm} = \frac{3}{8} \sigma_{\rm Th} \frac{c\delta t}{\alpha-1}
\Gamma_b n_{\gamma}^{'2} \left( \frac{2\epsilon^2_{\gamma, {\rm pk}}}
{m_e^2c^4 \Gamma_b^2} \right)^{\alpha-1}.
\label{tot-pairdensity-simple} \ea
The total number density of target photons in equation
(\ref{co-phot-den}) increases by a factor $\sim {\rm
ln}(\epsilon_{\gamma,{\rm pk}}/ \epsilon_{\gamma,{\rm ssa}})$ as the
target photon energy decreases from $\epsilon_{\gamma,{\rm pk}}$ to
$\epsilon_{\gamma,{\rm ssa}}$ in case of $\alpha = 2$.  However, one
needs incident photons of energy much higher than
$\epsilon_{\gamma,{\rm pk}}$ to produce $e^{\pm}$ pairs with photons
at $\epsilon_{\gamma,{\rm ssa}}$, and the incident number decreases
$\propto E_{\gamma}^{'-\alpha}$. Thus equation
(\ref{tot-pairdensity-simple}) is estimated for targets which are at
threshold with incident photons of energy $\geq \epsilon_{\gamma,{\rm
pk}}$. We can express the secondary pair density as the ratio to the
initial electron density in the fireball,
\ba \frac{n'_{\pm}}{n'_e} &=& \frac{3}{128\pi} \frac{\sigma_{\rm Th}}
{\alpha-1} \frac{L_{\gamma,{\rm iso}} \varepsilon_e m_p} {\Gamma_b^3
\delta t \epsilon_{\gamma,{\rm pk}}^2} \left(
\frac{2\epsilon^2_{\gamma, {\rm pk}}} {m_e^2c^4 \Gamma_b^2}
\right)^{\alpha-1} \nonumber \\ &\approx& \cases{ 786 L_{\gamma,52}
\varepsilon_{e,-1} \delta t_{-2}^{-1} \Gamma_{b,2.5}^{-5};~\alpha = 2
\cr 2.3 L_{\gamma,52} \varepsilon_{e,-1} \delta t_{-2}^{-1}
\epsilon_{\gamma,2.7} \Gamma_{b,2.5}^{-6};~\alpha = 2.5 \cr 0.008
L_{\gamma,52} \varepsilon_{e,-1} \delta t_{-2}^{-1}
\epsilon_{\gamma,2.7}^2 \Gamma_{b,2.5}^{-7};~\alpha = 3 }
\label{pair-ratio} \ea
using equation (\ref{co-part-den}).

Shock accelerated high energy electrons may also produce $e^{\pm}$
pairs interacting with synchrotron photons ($e\gamma \rightarrow e
e^{\pm}$). (The cross-section for this process is given by equation
(\ref{BH-cross}) with the replacement $E'_{\gamma}/m_e \rightarrow
\gamma'_e E'_{\gamma}/m_e$).  However, the number of secondary pairs
produced in this case is negligible compared to the initial lepton
number density because of the lower number density of high energy
incident electrons.

The secondary pairs can annihilate with themselves or with the leptons
originally carried in the fireball, $e^+e^- \rightarrow \gamma\gamma$,
if the number density of pairs increases substantially.  The
cross-section for this process is \citep{jauch}
\ba \sigma_{e^{\pm}} = \frac{3}{8} \frac{\sigma_{\rm Th}}{\gamma'_e
(\gamma'_e+1)} \left[ \left( \gamma'_e + 4 + \frac{1}{\gamma'_e}
\right) {\rm ln} \left( \gamma'_e + \sqrt{\gamma^{'2}_e -1} \right) -
\left( \gamma'_e + 3 \right) \right].
\label{pair-ann} \ea
Because of a decreasing cross-section with energy, low energy
$e^{\pm}$ pairs annihilate faster. The annihilation time for an
incident electron can be estimated as $t'_{\rm ann} = 1/(c
\sigma_{e^{\pm}} n'_{\pm})$ in the comoving frame. However,
synchrotron cooling is even faster and the ratio of pair annihilation
time and synchrotron cooling time, for example at $\gamma'_e = 5$
(where the electron is still relativistic and synchrotron loss formula
applies), $\Gamma_b = 10^{2.5}$, $\delta t = 0.01$ s and in $\alpha
=2$ case in equation (\ref{pair-ratio}), is $t'_{e,{\rm syn}}/t'_{\rm
ann} \approx 0.1$ using equations (\ref{syn-cooltime}) and
(\ref{pair-ann}). This ratio decreases at higher energy. Hence
electrons lose most of their energy to synchrotron photons within a
dynamical time $\delta t \Gamma_b$ in the comoving frame. As a result,
secondary pairs produced in the fireball cool down to nonrelativistic
values and drift along roughly with the bulk Lorentz factor
$\Gamma_b$.  We consider here situations where the secondary pair
density, equation (\ref{pair-ratio}), is at most comparable to the
original fireball lepton density, in which case the scattering depth
is not substantially changed, and any photons from annihilation of
cold pairs lead to higher generation pairs whose number similarly does
not affect the scattering optical depth.  We note that in the Thomson
limit, inverse Compton (IC) cooling time is of the same order as the
synchrotron cooling time $t'_{e,{\rm IC}}/t'_{e,{\rm syn}} \approx
\varepsilon_e/\varepsilon_B$ with an increasing value at higher energy
and in the Klein-Nishina limit.

\section{High energy photon spectrum and interaction with the diffuse
background}

\begin{figure}
\plotone{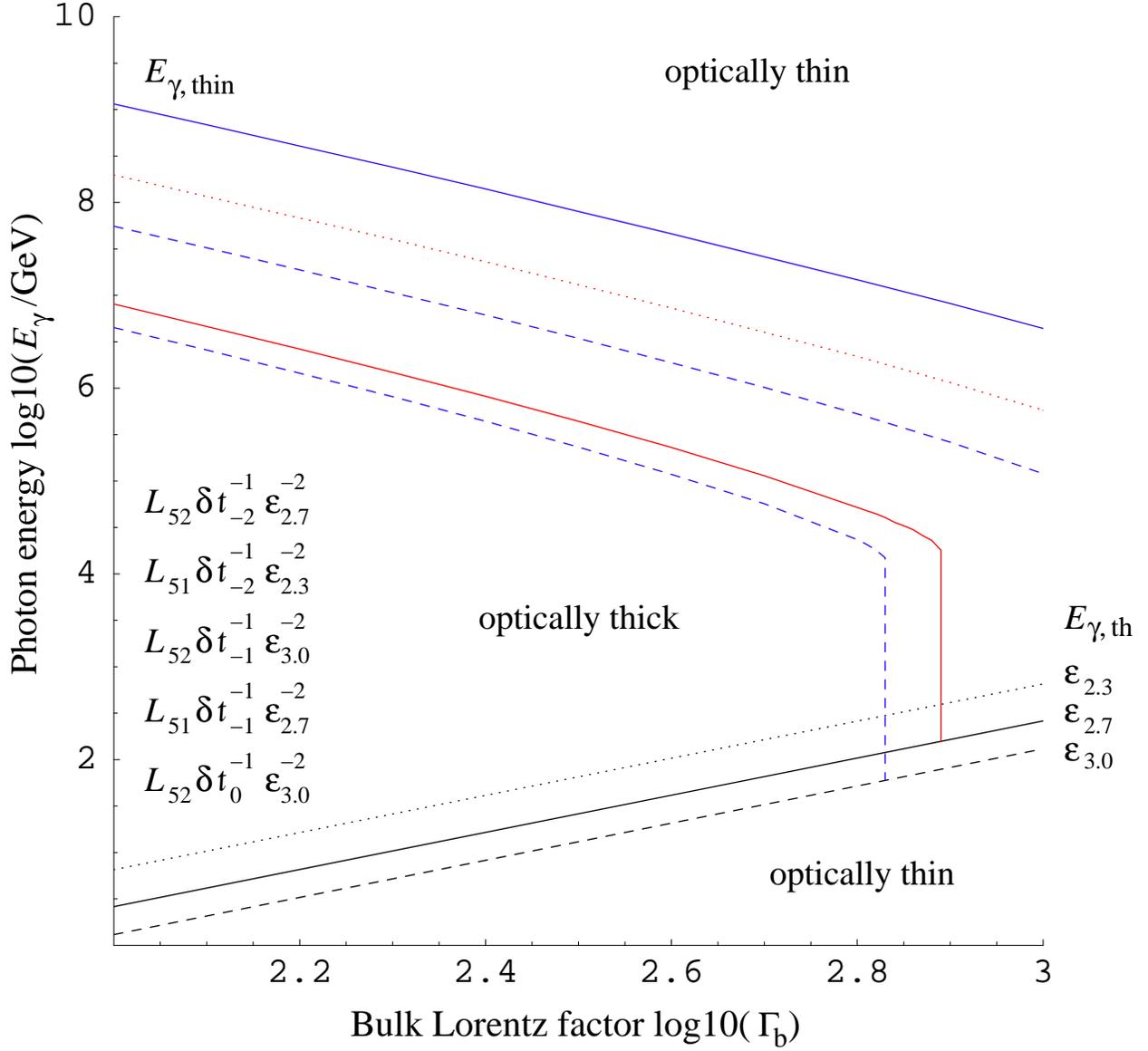} \caption{Cartoon of allowed (optically thin) and
forbidden (optically thick) photon energy range, in the observer's
frame, from the GRB fireball as a function of the bulk Lorentz factor
($\Gamma_b$). Photons below $E_{\gamma,{\rm th}}$ and above
$E_{\gamma,{\rm thin}}$ may escape. We have plotted the curves for 2
different $\gamma$-ray luminosity $10^{52}L_{52}$, $10^{51}L_{51}$; 4
different variability time $1 \delta t_{0}$ s, $0.1 \delta t_{-1}$ s,
$0.01 \delta t_{-2}$ s, $0.001 \delta t_{-3}$ s; and 3 different peak
synchrotron photon energy $200 \varepsilon_{2.3}$ keV, $500
\varepsilon_{2.7}$ keV, $1000 \varepsilon_{3.0}$ keV. However, we have
kept the synchrotron self-absorption energy to be fixed at 10 keV in
all cases. The parameter choices are shown for curves from top to
bottom. \label{wphase}}
\end{figure}

Using the phenomenological Band function spectrum for the intrinsic
high energy photon flux from the GRB, the spectrum that would reach
the observer after taking into account in-source $\gamma\gamma$
attenuation at a luminosity distance $D_L$ is
\ba \frac{d^2N_{\gamma}}{d \epsilon_{\gamma} dt} = \frac{(\alpha -1)
L_{\gamma,{\rm iso}}}{4 \pi D_L^2 \epsilon_{\gamma,{\rm pk}}^2} \left(
\frac{\epsilon_{\gamma}}{\epsilon_{\gamma,{\rm pk}}} \right)^{-\alpha}
e^{-\tau_{\gamma\gamma,{\rm grb}}(\epsilon_{\gamma})}; \;\;
\epsilon_{\gamma} \gtrsim \epsilon_{\gamma,{\rm pk}}
\label{prompt-flux} \ea
where $\tau_{\gamma\gamma,{\rm grb}}(\epsilon_{\gamma})$ is the
$e^{\pm}$ pair production optical depth (see in Figure
\ref{wlimits52}) in the GRB fireball. The high optical depth in the
fireball, however, reduces substantially the ultra-high energy photon
flux above the energy $E_{\gamma,{\rm pk,th}}$ given in equation
(\ref{egampair-thresh}) and below the energy $E_{\gamma,{\rm thin}}$
given in equation (\ref{egammax-thin}), while for some parameters the
GRB is in the region of Figure \ref{wphase} where the escaping
spectrum is optically thin at all energies. Here we consider high
energy photon emission from the GRB fireball for two cases. One is an
example where the escaping spectrum is optically thin at all energies
(e.g. $L_{\gamma,{\rm iso}} = 10^{52}$ erg/s, $\delta t = 1$ s,
$\epsilon_{\gamma,{\rm pk}} = 1$ MeV and $\Gamma_b = 10^{2.9}$ in
Figure \ref{wphase}).  The other example is a case when the fireball
is opaque to photons in an energy range between $E_{\gamma,{\rm
pk,th}}$ and $E_{\gamma,{\rm thin}}$ (e.g. $L_{\gamma,{\rm iso}} =
10^{52}$ erg/s, $\delta t = 0.1$ s and $\epsilon_{\gamma,{\rm pk}} =
0.5$ MeV and $\Gamma_b = 10^{2.5}$).

For typical GRB locations at redshift $z=1$, most high energy photons
above $\sim 70$ GeV produce $e^{\pm}$ pairs \citep{stecker98},
reaching an energy up to $\sim m_e\gamma_{e, {\rm max}}/2$, due to
$\gamma\gamma$ interactions with cosmic infra-red background (CIB)
photons (in the $\sim 100$ GeV-TeV range), and cosmic microwave
background (CMB) photons (in the PeV range). These high energy pairs,
in turn, cause delayed high energy photon emission by inverse Compton
(IC) scattering of CMB photons \citep{dai02a}. Here we follow the
treatment of \citet{dai02b} to calculate the delayed high energy
photon spectrum in the GeV-TeV range.

The corresponding time integrated flux of electrons (and positrons)
from high energy GRB photons interacting with CIB and CMB photons is
\ba \frac{dN_e}{d\gamma_e} &=& \frac{L_{\gamma,{\rm iso}} \delta t}
{2\pi D_L^2} \frac{\alpha -1}{(2m_e)^{\alpha -1}}
\frac{\epsilon_{\gamma,{\rm pk}}^{\alpha -2}}{\gamma_e^{\alpha}}
e^{-\tau_{\gamma\gamma,{\rm grb}}(2\gamma_e m_e)} \nonumber \\ &&
\times \left[ 1- e^{-\tau_{\gamma\gamma,{\rm bkg}}(2\gamma_e m_e)}
\right] \Theta \left( \frac{m_e}{4 \epsilon_{\gamma,{\rm cib;cmb}}}
\lesssim \gamma_e \lesssim \frac{\gamma_{e,{\rm max}}}{2} \right),
\label{cib-leptons} \ea
where $\Theta$ is a step function. We have assumed that each electron
and positron of the $e^{\pm}$ pair share 1/2 the photon energy,
i.e. $\gamma_e = \epsilon_{\gamma}/2m_e$. We calculate the pair
production optical depth with background photons (CIB and CMB) at
threshold ($2\epsilon_{\gamma} \epsilon_{\gamma,{\rm cib;cmb}} =
m_e^2$) as
\ba \tau_{\gamma\gamma,{\rm bkg}} (\epsilon_{\gamma}) = {\rm max} \{
n_{\rm cib} ( m_e^2/2\epsilon_{\gamma,{\rm cib}}), n_{\rm cmb}
(m_e^2/2\epsilon_{\gamma,{\rm cmb}}) \} \times (3/8) \sigma_{\rm Th}
D_L
\label{cib-opt} \ea
The total number density of CIB photons is fitted from the intensity
$\mathcal I$ (W-m$^{-2}$ sr$^{-1}$) given in \citet{malkan01} as
\ba n_{\rm cib} (\epsilon_{\gamma,{\rm cib}}) = \frac{4\pi {\mathcal
I} (1+z)^3}{c \epsilon_{\gamma,{\rm cib}}}. \ea
The CMB photons, on the other hand, have a black body distribution
peaking at an energy of $2.7(1+z)$ K and a total volume number density
$\propto (1+z)^3$.

\begin{figure}
\plotone{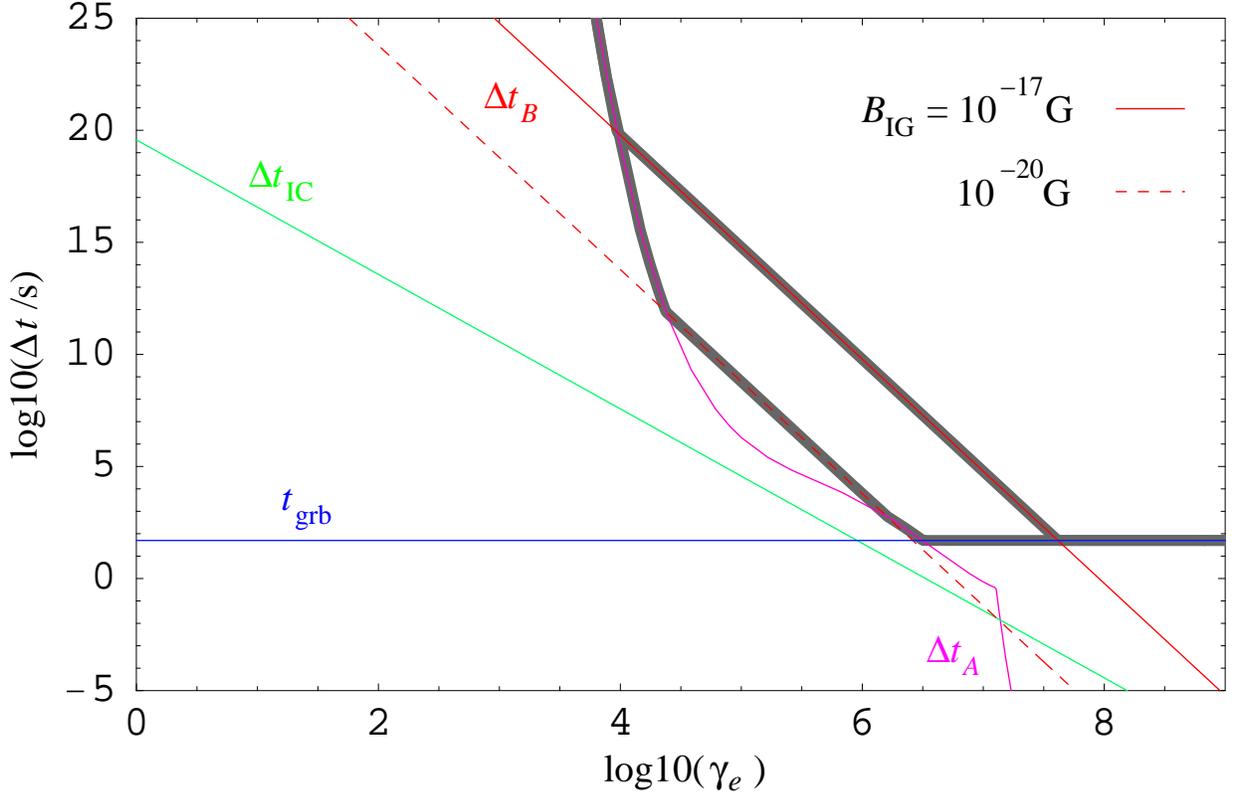} \caption{Different time scales ($\Delta t_{\rm IC}$,
$\Delta t_{A}$, $\Delta t_{B}$, and $t_{\rm grb}$), involved in
calculatng the delayed secondary IC spectrum using equation
(\ref{tot-cib-spect}), as a function of secondary $e^{\pm}$ pair
energy ($\gamma_e$). The thick gray lines correspond to $\Delta t
(\gamma_e) = {\rm max} \{ \Delta t_{\rm IC}, \Delta t_{A}, \Delta
t_{B}, \delta t \}$ for two different $\Delta t_B$, denoted by solid
and dashed lines, corresponding to inter-galactic magnetic field
strength ($B_{\rm IG}$) of $10^{-17}$ and $10^{-20}$ G
respectively. We have used a GRB duration of $t_{\rm grb} = 50$ s.
All times are plotted for a fixed luminosity distance of $D_L =
10^{28}$ cm. \label{times}}
\end{figure}

The secondary electrons in equation (\ref{cib-leptons}) are cooled by
IC scattering on CMB photons on a timescale $t_{\rm IC}(\gamma_e) =
3m_e c/(4 \gamma_e \sigma_{\rm Th} u_{\rm cmb}) \approx 7.3 \times
10^{13} (\gamma_e/10^6)^{-1}$ s in the local rest frame
\citep{dai02b}, where $u_{\rm cmb}$ is the local black body CMB photon
energy density. The total integrated, first generation, IC photon
spectrum in the observer's frame, convolving the electron spectrum
with the IC spectrum from a single electron in the Thomson limit
\citep{blumenthal70}, is then
\ba \frac{d^2 N_{\rm delayed~IC}}{dt dE_{\gamma}} &=& \int \int
d\gamma_e \frac{dN_e}{d\gamma_e} \frac{d^2 N_{\gamma}}{dt dE_{\gamma}}
\nonumber \\ &=& \frac{3 \sigma_{\rm Th}c}{32 \pi D_L^2}
\frac{(\alpha-1) \epsilon_{\gamma,{\rm pk}}^{\alpha -2}}{(2
m_e)^{\alpha -1}} L_{\gamma,{\rm iso}} t_{\rm grb}~ {\rm exp}
\left[-\tau_{\gamma\gamma, {\rm bkg}}(E_{\gamma}) \right] \nonumber \\
&& \times \int d\gamma_e \gamma_e^{-(\alpha+4)}~{\rm exp} \left[
-\tau_{\gamma\gamma,{\rm grb}}(2\gamma_e m_e) \right] \left( 1- {\rm
exp} \left[ -\tau_{\gamma\gamma,{\rm bkg}}(2m_e \gamma_e) \right]
\right) \frac{t_{\rm IC}(\gamma_e)}{\Delta t(\gamma_e)} \nonumber \\
&& \times \int d\epsilon_{\gamma,{\rm cmb}} \frac{n_{\rm cmb}
(\epsilon_{\gamma,{\rm cmb}})}{\epsilon_{\gamma,{\rm cmb}}^2} \left[ 2
E_{\gamma} {\rm ln} \frac{E_{\gamma}}{4 \gamma_e^2
\epsilon_{\gamma,{\rm cmb}}} + E_{\gamma} + 4 \gamma_e^2
\epsilon_{\gamma,{\rm cmb}} - \frac{E_{\gamma}^2}{2 \gamma_e^2
\epsilon_{\gamma,{\rm cmb}}} \right].
\label{tot-cib-spect} \ea
Here the integration over $\gamma_e$ ranges from $\gamma_{e,{\rm min}}
= {\rm max} \{ m_e/(4 \epsilon_{\gamma,{\rm cmb}}),
\sqrt{E_{\gamma}/\epsilon_{\gamma,{\rm cmb}}}/2 \}$ to $\sim
\gamma_{e,{\rm max}}/2$ where $\gamma_{e,{\rm max}}$ is the maximum
electron energy in equation (\ref{max-eboost}). The maximum time scale
for delayed emission is $\Delta t (\gamma_e) = {\rm max} \{ \Delta
t_{\rm IC}, \Delta t_{A}, \Delta t_{B}, t_{\rm grb} \}$, where $\Delta
t_{\rm IC}(\gamma_e) = t_{\rm IC}/(2\gamma_e^2)$ is the IC cooling
time; $\Delta t_{A}(\gamma_e) = {\lambda}_{\gamma\gamma,{\rm
bkg}}/(2\gamma_e^2 c)$ is the angular spreading time; $\Delta
t_{B}(\gamma_e) = t_{\rm IC} \theta_{B}^2/2$ is the delay time due to
magnetic deflection, all evaluated in the observer's frame; and
${\lambda}_{\gamma\gamma, {\rm bkg}} = ( {\rm max} \{ n_{\rm cib},
n_{\rm cmb} \} (3/8) \sigma_{\rm Th})^{-1}$ is the pair production
mean free path against background photons. For $10^{-20}B_{{\rm
IG},-20}$ G inter galactic (IG) magnetic field, the deflection angle,
for $\theta_{B} \ll 1$, is $\theta_{B} \approx \lambda_{e}/R_{\rm g}
\approx 1.3 \times 10^{-5} (\gamma_e/10^6)^{-2} (B_{{\rm IG},-20})^2$,
where $R_{\rm g} = \gamma_e m_e c^2/(eB)$ is the Larmor gyroradius of
the electron. We have plotted different time scales as a function of
secondary pair Lorentz factor ($\gamma_e$) in Figure \ref{times} for
$t_{\rm grb} = 50$ s and several IG magnetic field values, at $D_L =
10^{28}$ cm corresponding to redshift $z \simeq 1$ in an $\Omega_m =
0.3$, $\Omega_\Lambda =0.7$ and $H_o = 75$ km s$^{-1}$ Mpc$^{-1}$
cosmology.

\begin{figure}
\plotone{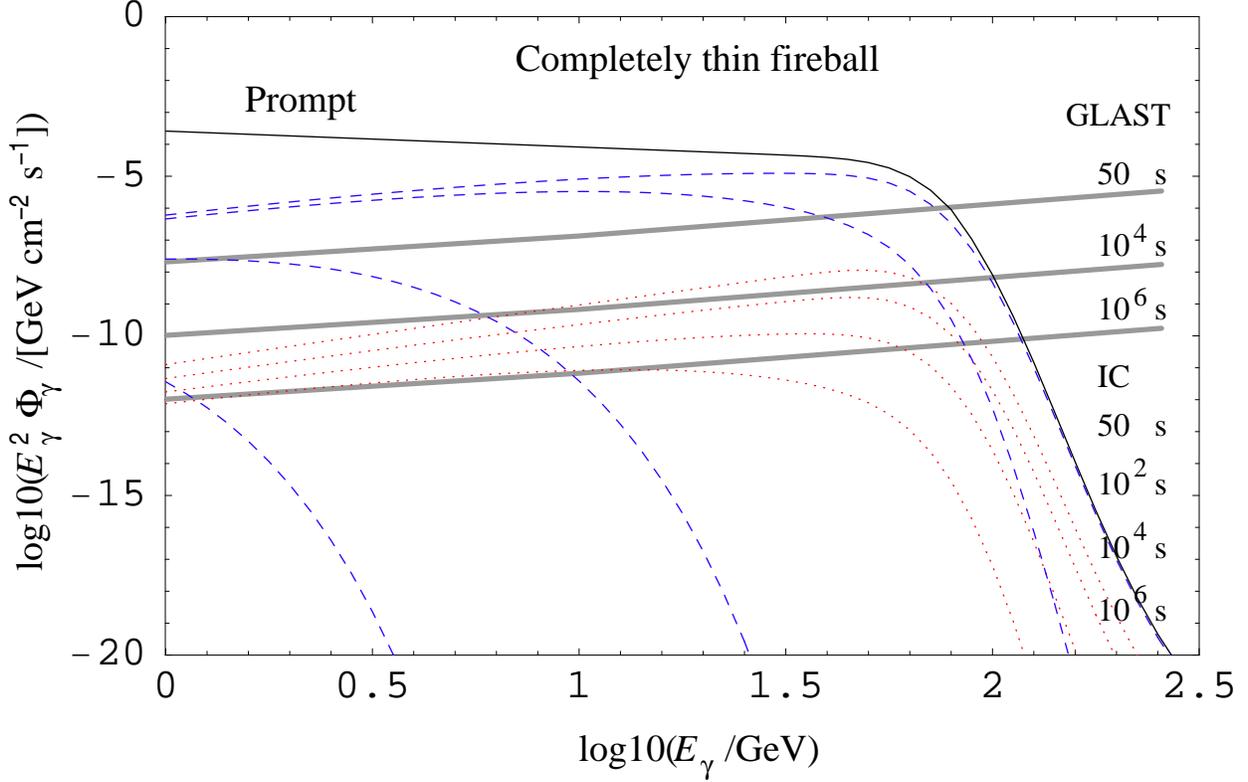} \caption{High energy photon spectrum ($E_{\gamma}^2
\Phi_{\gamma} = E_{\gamma}^2 d^2N_{\gamma}/dt dE_{\gamma}$) from a GRB
where the fireball is $\gamma\gamma$ optically thin to photons of all
energy. The primary spectrum (full line) is due internal shocks, the
secondary spectrum (dashed and dotted) is due to $\gamma\gamma$
interactions and IC scattering in the external environment. The
parameters are $\delta t = 1$ s, $\epsilon_{\gamma,{\rm pk}} = 1$ MeV,
$\Gamma_b = 10^{2.9}$, isotropic-equivalent luminosity $10^{52}$
erg/s, spectral index $\alpha = 2.5$, located at redshift $z=1$.  The
delayed spectra are calculated for two different intergalactic
magnetic field values, $B_{\rm IG} = 10^{-20}$ and $10^{-17}$ G,
denoted by the dashed and the dotted curves respectively. The duration
of the delayed secondary emission (from top to bottom) spectra are 50
s, $10^2$ s, $10^4$ s and $10^6$ s. The GLAST sensitivity (thick gray
lines) for 50 s, $10^4$ s and $10^6$ s integration time is also
plotted for comparison. \label{spectrum1}}
\end{figure}

\begin{figure}
\plotone{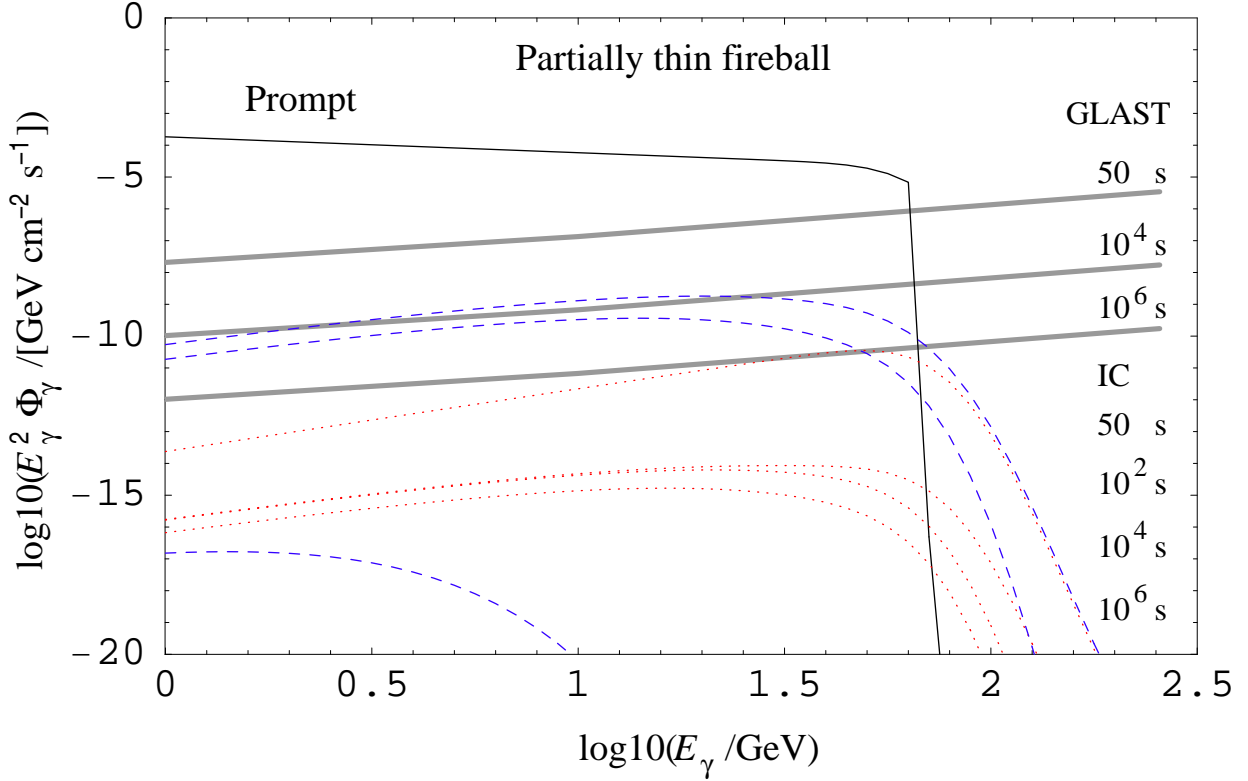} \caption{High energy photon spectrum ($E_{\gamma}^2
\Phi_{\gamma} = E_{\gamma}^2 d^2N_{\gamma}/dt dE_{\gamma}$) from a GRB
where the fireball is optically thick due to $\gamma\gamma$ against
some its own photons over a certain energy band. The parameters are
$\delta t = 0.1$ s, $\epsilon_{\gamma,{\rm pk}} = 0.5$ MeV, $\Gamma_b
= 10^{2.7}$, isotropic-equivalent luminosity $10^{52}$ erg/s, spectral
index $\alpha = 2.5$, and redshift $z=1$. Plotted are the prompt
(solid curve) and delayed spectra (dashed and dotted curves).  The
delayed spectra are for two different intergalactic magnetic fields of
$B_{\rm IG} = 10^{-20}$ and $10^{-17}$ G, denoted by the dashed and
the dotted curves respectively. The duration of delayed emission (from
top to bottom) spectra are 50 s, $10^2$ s, $10^4$ s and $10^6$
s. GLAST sensitivities (thick gray lines) for 50 s, $10^4$ s and
$10^6$ s integration time are also plotted for
comparison. \label{spectrum2}}
\end{figure}

We used two parameter sets ($L_{\gamma,{\rm iso}} = 10^{52}$ erg/s,
$\delta t = 1$ s, $\epsilon_{\gamma,{\rm pk}} = 1$ MeV and $\Gamma_b =
10^{2.9}$) and ($L_{\gamma,{\rm iso}} = 10^{52}$ erg/s, $\delta t =
0.1$ s and $\epsilon_{\gamma,{\rm pk}} = 0.5$ MeV and $\Gamma_b =
10^{2.7}$). These correspond to a case when the primary spectrum
escaping from the GRB fireball is optically thin to photons of all
energes, and a case when the primary fireball escaping spectrum is
optically thick to $\gamma\gamma$ against its own photons in a band of
photon energies, shown in Figures \ref{spectrum1} and \ref{spectrum2}
respectively.  The prompt primary spectra, using equation
(\ref{prompt-flux}) multiplied by $\exp[-\tau_{\gamma\gamma,{\rm
bkg}}(\epsilon_\gamma)]$ to allow for attenuation in the background
radiation fields, are plotted as the dark solid curves in both
figures. We have also plotted the first generation delayed spectra
from IC scattering of secondary leptons on the CMB, by numerically
integrating equation (\ref{tot-cib-spect}) for two different IGM
magnetic field strengths ($B_{\rm IG}$) of $10^{-20}$ G and $10^{-17}$
G. For the same two sets of parameters above these are given by the
dashed and dotted curves respectively.  The delayed spectra for 50 s
duration (topmost dashed and dotted curves) corresponds to the GRB
duration ($t_{\rm grb}$) and we have calculated them using the full
range of secondary pair energy $\gamma_e = \epsilon_{\gamma}/2m_e$,
created with CMB, allowed by the maximum primary photon energy
$\epsilon_{\gamma} \approx \gamma'_{e, {\rm max}} m_e \Gamma_b$ from
equation (\ref{max-eboost}). We have also calculated the delayed
spectra for $10^2$, $10^4$ and $10^6$ s duration by numerically
integrating equation (\ref{tot-cib-spect}) up to a maximum $\gamma_e$
corresponding to that time from Figure \ref{times}.

\section{Discussion}

We have calculated the spectrum and time dependence of the high energy
($\gtrsim$ GeV) spectrum emerging from GRB internal shocks as a
function of the luminosity, time variability and the bulk Lorentz
factor, using exact cross sections.  For typical bursts parameters the
internal shock spectrum cuts off above a threshold energy
$E_{\gamma,{\rm th}}\sim 10-100$ GeV due to $\gamma\gamma$ iteractions
within the shock region.  There is a qualitative difference with the
results of \citet{baring94}, who discussed the $\gamma\gamma$
absorption in external shocks, and who did not consider the secondary
spectra from interactions with background radiation. More recently
\citet{wang04} also discussed the $\gamma\gamma$ absorption in
external shocks, including interactions against background
photons. The main difference with our work is due to the much lower
$\gamma\gamma$ optical depth from the lower photon density at the
radii of external shocks, which are much larger than the radii of
internal shocks considered here.  Internal shock $\gamma\gamma$
absorption using approximate cross sections was discussed by
\citet{pilla98} and \citet{lithwick01} without interactions against
external background photons, and by \citet{dai02a} including such
interactions.  These authors did not consider the effects of
synchrotron self-absorption.  The main qualitative difference between
the emerging internal shock spectra of these authors and our work
arises from our use of accurate cross sections, and more importantly,
our inclusion of synchrotron self-absorption.  The cut-offs in the
emerging spectra that we obtain are compatible with more detailed
numerical calculations of the self-consistent pair creation and
annihilation spectrum in internal shocks of \citet{peer03}, without
inclusion of interactions against external background photons. The
main difference between their final spectra and ours is due to their
using an ab-initio spectrum instead of a phenomenological Band
spectrum as here, and to the additional effects of our inclusion here
of interactions againts external photons, leading to a delayed
secondary spectrum.

A new feature discussed in this paper is that the spectra that emerge
from the GRB internal shock region shows, besides the expected high
energy cutoff due to pair-production $E_{\gamma,{\rm th}}$, also a
re-emergence of the spectrum at a higher energy $E_{\gamma,{\rm thin}}
> E_{\gamma,{\rm th}}$ where the shock becomes optically thin again to
$\gamma\gamma$ interactions.  This is due to our inclusion of
synchrotron self-absorption, which reduces the internal photon target
spectral density causing absorption of the highest energy emerging
photons. For high enough bulk Lorentz factors (typically $\Gamma_b
\gtrsim 800$), the fireball is optically thin to internal
$\gamma\gamma$ interactions at all energies, and there is no internal
$\gamma\gamma$ cut-off band in the emerging spectrum.

The emerging GRB spectrum is modified, on its way to the observer, by
interactions with the diffuse infrared background (CIB) and the
diffuse microwave background (CMB). Here, we have used the GRB
internal shock emerging spectra including synchrotron self-absorption
as the source of input photons, and calculate their interaction with a
CIB spectrum redshifted to $z\sim 1$, based on that of
\citet{malkan01}.  We then calculated the secondary GRB spectrum from
the resulting secondary pairs which are up-scattered on CMB photons.
We used two different escaping GRB primary spectra, one with an
internal $\gamma\gamma$ cutoff band (for a typical burst bulk Lorentz
factor $\Gamma_b\sim 300$) and one without an internal $\gamma\gamma$
cutoff band (for a burst with a larger $\Gamma_b \sim 800$). We used
two different isotropic-equivalent luminosities, and different
intergalactic magnetic field strengths ranging between $B_{\rm IG}=
10^{-20}$ G and $10^{-17}$ G. This leads to a secondary spectrum,
mainly in the 1-100 GeV range, in addition to and delayed respect to
the prompt unabsorbed internal shock spectrum.  We also calculated the
further interaction of the secondary gamma-ray spectrum with the CIB,
which gives a similarly attenuated spectral shape as that of the
primary spectrum. We did not consider the delayed spectrum of the
tertiary and higher order pairs, since the delay time scales are
longer and the fluxes of these components is much lower.

A different delayed GeV component can arise in the afterglow phase,
when the inverse Compton spectrum peak energy in the external shock
region sweeps across the GeV band \citep{mr94, dermer00, se01, zm01}.
This delayed GeV emission, from IC upscattering on external shock
electrons, is predicted to be detectable by GLAST at $z \sim 1$
\citep{zm01}, as long as the shock parameters are such that the IC
component is prominent (which is the case for the typical parameters
inferred from broadband modeling of some well-studied bursts
\citep{wijers99, panaitescu02, yost03}).  The IC duration can be up to
a few hours, which could overlap with the earlier part of the delayed
spectra discussed here.  The temporal evolution of the IC spectral
component is such that it hardens with time initially, but softens
with time after the IC peak crosses the band. The delayed GeV
component discussed in this paper has a rather different temporal
evolution behavior. Since $\Delta t (\gamma_e)$ is essentially
anti-correlated with $\gamma_e$ (see Figure 3), the hard spectrum (due
to the more energetic electrons) lasts a shorter timescale than the
soft spectrum. Also the harder photons arrive earlier than the softer
photons. Throughout the whole delayed phase, the hardest photons
emerge first, followed by progressively softer photons. Later the
hardest photons drop out first from the spectrum.  The delayed
spectrum always progressively softens with time.  This distinct
spectral evolution behavior can be used to distinguish the delayed
emission from IC scattering in external shocks from the delayed
emission of secondary spectra due to pair-production and IC scattering
on the diffuse background discussed here. The secondary external
spectrum from such an IC spectrum, however, should be similar to what
we have discussed here, on timescales longer than a few hours, since
the absorbed energy undergoes similar reprocessing.

The time delays and the flux level of the secondary spectra are
sensitive to the intergalactic magnetic field strength. The best
prospects for detection of such secondary delayed GeV spectra occurs
for bursts with bulk Lorentz factors $\gtrsim 800$ and intergalactic
magnetic fields $\lesssim 10^{-17}$ G, which might be typical of
intergalactic void regions. Since the mean-free-path of the first
generation high energy photons against $\gamma\gamma$ absorption
typically exceeds several to tens of Mpc \citep{coppi97, malkan01}, it
is plausible that a substantial fraction of the pair formation will
occur in void regions near the burst, even if the latter occured in a
cluster galaxy.  The corresponding time delays are in the range 10-100
s and the fluxes are within the sensitivity range of GLAST, and at the
high energy end, also of VERITAS, HEGRA, and other ACTs.  A detection
of this secondary delayed GeV spectrum in GRB would provide a valuable
diagnostic for the IGM magnetic field strength, as well as the bulk
Lorentz factor, and would provide valuable clues for the typical size
of the shocks responsible for the radiation.

\acknowledgements

This research is supported in part by NSF AST 0098416, NASA NAG5-13286
and the Monell Foundation.

\clearpage

\clearpage

\clearpage

\clearpage

\clearpage

\end{document}